\documentstyle[prd,floats,aps]{revtex}

\input psfig
\pssilent

\begin{document}

\title{Close limit evolution of Kerr-Schild type initial data for binary
black holes}

\author{Olivier Sarbach}
\address{Center for Gravitational Physics and Geometry,
Department of Physics,\\
The Pennsylvania State University, 104 Davey Lab, University Park, PA 16802.}

\author{Manuel Tiglio}
\address{Center for Gravitational Physics and Geometry,
Department of Physics and Department of Astronomy and Astrophysics, \\
The Pennsylvania State
University, 525 Davey Lab, University Park, PA 16802.}

\author{Jorge Pullin}
\address{Department of Physics and Astronomy, Louisiana State University, 202
Nicholson Hall,  Baton Rouge, LA 70803-4001.}
\maketitle

\begin{abstract}
We evolve the binary black hole initial data family proposed by Bishop
{\em et al.} in the limit in which the black holes are close to each
other. We present an exact solution of the linearized initial value
problem based on their proposal and make use of a recently introduced
generalized formalism for studying perturbations of Schwarzschild black holes
in arbitrary coordinates to perform the evolution.  We clarify the
meaning of the free parameters of the initial data family through the results
for the radiated energy and waveforms from the black hole collision.
\end{abstract}

\section{Introduction}

Collisions of binary black holes are expected to be one of the primary
sources of gravitational radiation to be detected by interferometric
gravitational wave detectors. Given the non-symmetric, time-dependent
nature of the problem, the only realistic hope of modeling a collision
is via numerical simulations. Unfortunately, given computer
limitations, it is expected that in the near future most evolutions
will have to start with the black holes quite close to each
other. This brings to the forefront the problem of specifying initial
data for the binary black hole collisions. Ideally, one would like to
have initial data representing ``astrophysically relevant''
situations, that is, resembling the situation the two black holes
would be in when they are in a realistic collision at the given
separation. Unfortunately, providing realistic data is tantamount to
solving the evolution problem. Since this cannot be done, one is left
with generating families of initial data based on mathematical or
computational convenience. Initial attempts to provide initial data
concentrated on solutions to the initial value problem that were
conformally flat \cite{C}. Conformally flat spatial metrics simplify
considerably the constraint equations but suffer from drawbacks, most
notably the inability to incorporate Kerr black holes, which are not
known to have conformally flat sections (for a perturbative proof of
non-existence, see \cite{price}).  Recently, attention has been drawn
to the construction of initial data based on the Kerr--Schild form of
the Schwarzschild and Kerr metrics. These constructions have several
attractive properties: the slices are horizon-penetrating, which makes
them suitable for the application of the ``excision'' technique for
evolving black holes, and they can naturally incorporate boosted and
spinning black holes \cite{fd}.

There have been two different
proposals to use Kerr--Schild coordinates for binary black holes. In the proposal of Huq,
Matzner and Shoemaker \cite{HMS},
two black holes individually in Kerr--Schild form were
superposed. In the proposal of Bishop {\em et al.} \cite{BIMW}
the superposition was
carried out in a way that the resulting superposed metric was in
Kerr--Schild form. This latter proposal has the property that in the
``close limit'' in which the separation of the holes is small, the
metric is given by a distorted Kerr-Schild black hole.

In this paper we will consider the evolution of the Bishop {\em et al.}
family of initial data in the close limit, by treating the
space-time as a small perturbation of a non-rotating Kerr--Schild black hole.
The evolution will be carried out using a recently introduced perturbative
formalism that allows to evolve Schwarzschild black holes in arbitrary
spherically symmetric coordinates \cite{ST}.
We will present an explicit solution to the initial value
problem posed by Bishop {\em et al.} in the close limit and use it to
compute the radiated energy in the black hole collision. This in turn
will help use clarify the meaning of free parameters that appear in
these families of data.

\section{Kerr-Schild initial data}

In their paper, Bishop {\em et al.} \cite{BIMW} assume that
at the initial slice, the three metric and the extrinsic
curvature are of Kerr-Schild (KS) type.
The KS space-time metric is defined by
\begin{displaymath}
g_{\mu\nu} = \eta_{\mu\nu} - 2V k_\mu k_\nu\, ,
\end{displaymath}
where $k_\mu$ is null.
The ``background'' metric $\eta_{\mu\nu}$ is taken to be the
Minkowski metric with coordinates $(t,\underline{x}) = (t,x^i)$ such that
$\eta_{tt} = -1$, $\eta_{ti} = 0$ and $\eta_{ij} = \delta_{ij}$.
The null vector $k_\mu$ satisfies $k_t = -1$ and
$k^i k_i = 1$, where $k^i = \delta^{ij} k_j$.
As an example, for the Schwarzschild geometry,
\begin{displaymath}
V = -\frac{M}{r}\, ,\;\;\;
k_i dx^i = -dr.
\end{displaymath}
For a KS metric, the three metric and extrinsic curvature
with respect to a slice $t = \mbox{const.}$ are
\begin{eqnarray}
\bar{g}_{ij} &=& \delta_{ij} - 2V k_i k_j,
\label{Eq-KSgK1}\\
K_{ij} &=& -\frac{1}{\alpha}\partial_t \left( V k_i k_j \right)
 + 2\alpha \left[ V k^s \nabla_s \left( V k_i k_j \right) -
 \nabla_{(i} \left( V k_{j)} \right) \right],
\label{Eq-KSgK2}
\end{eqnarray}
where $\alpha = (1 - 2V)^{-1/2}$ is the lapse and where
$\nabla$ refers to the flat metric $\delta_{ij}\,$.
Bishop {\em et al.}'s solution procedure consists in inserting
(\ref{Eq-KSgK1},\ref{Eq-KSgK2}) into the constraint equations
and to solve the resulting equations for $V$, $\dot{V} = \partial_t V$
and $\dot{k}_i = \partial_t k_i$, where $k_i$ is assumed to be given.
($k_i$ has only two independent components since $k^i k_i = 1$.)
Below, we review the discussion of these equation where a
particular ansatz is made for $k_i$ representing two nearby
non-rotating and non-spinning black holes.

\subsection{Two black hole data}

A single Schwarzschild black hole can be represented as
\begin{displaymath}
k_i = \frac{\nabla_i \phi}{|\nabla\phi|}\, , \;\;\;
|\nabla\phi|^2 = \delta^{ij} \nabla_i\phi\cdot\nabla_j\phi,
\end{displaymath}
with $\phi = 1/r$.
For two black holes, Bishop {\em et al.} make the ansatz
\begin{displaymath}
\phi(\underline{x}) = \frac{M_1}{ |\underline{x} -
\underline{x}_1 | } + \frac{M_2}{ |\underline{x} - \underline{x}_2 | }\, ,
\end{displaymath}
where $\underline{x}_{1,2}$ denote the position of the black hole $1,2$
which has mass $M_{1,2}$.
If the black holes are located at $\underline{x}_i = a_i(0,0,1)$, with
$a_1 > 0 > a_2$, $\phi$ may be expanded in a sum over multipoles:
\begin{equation}
\phi(r,\vartheta) = \sum\limits_{\ell=0}^{\infty}
\frac{ M_1 a_1^\ell + M_2 a_2^\ell}{r^{\ell+1}} P_\ell(\cos\vartheta),
\label{Eq-TwoBHExp}
\end{equation}
where $P_\ell$ denote standard Legendre polynomials and where
$(r,\vartheta,\varphi)$ are polar coordinates for $\underline{x}$.
It is important to note that the expansion (\ref{Eq-TwoBHExp})
is only valid for $r > \max\{a_1,-a_2\}$.

Defining the separation parameter
\begin{displaymath}
\varepsilon = \frac{a_1 - a_2}{M}\, ,
\end{displaymath}
where $M = M_1 + M_2$ is the total mass,
and imposing the center of mass condition $M_1 a_1 + M_2 a_2 = 0$,
the close limit of (\ref{Eq-TwoBHExp}) becomes
\begin{displaymath}
\phi = \frac{M}{r} + \varepsilon^2\frac{ M M_1 M_2}{r^3}
P_2(\cos\vartheta) + {\cal O}(\varepsilon^3/r^4).
\end{displaymath}
As a result, to first order in $\varepsilon^2$, $k_i$ is given by
\begin{equation}
k_r = 1, \;\;\;
k_A = \varepsilon^2\frac{M_1 M_2}{r} \hat{\nabla}_A P_2\, ,
\label{Eq-Exp1}
\end{equation}
where here and in the following, $A = \vartheta,\varphi$.
The remaining amplitudes are expanded according to
\begin{equation}
V = -\frac{M}{r} + \varepsilon^2 v(r) P_2\, , \;\;\;
\dot{V} = \varepsilon^2\dot{v}(r) P_2\, , \;\;\;
\dot{k}_A = \varepsilon^2\dot{k}(r) \hat{\nabla}_A P_2\, .
\label{Eq-Exp2}
\end{equation}
(In Bishop {\em et al.}'s notation: $\dot{v} = v_T$ and $\dot{k} = -k_T$.)
Introducing this into the constraint equations, and keeping only
terms of the order $\varepsilon^2$, one obtains the equations
\begin{eqnarray}
0 &=& -\dot{v} + \frac{3M}{r^2}\left( 1 + \frac{2M}{r} \right)\dot{k}
      - \frac{3}{r}v - \frac{6M M_1 M_2}{r^5}(r - M),\label{Eq-C1}\\
0 &=& -v' - \frac{4}{r} v + \frac{6M^2}{r^3} \dot{k}
      - \frac{6M M_1 M_2}{r^5}(r - M), \label{Eq-C2}\\
0 &=& - M \dot{k}' + v + \frac{2M}{r} \dot{k} + \frac{6M M_1 M_2}{r^3}\, .
\label{Eq-C3}
\end{eqnarray}
Here, a prime denotes differentiation with respect to $r$.
The system (\ref{Eq-C2},\ref{Eq-C3}) can be re-expressed as
a single second order equation. Introducing the dimensionless
quantities $x = r/M$ and $\mu = M_1 M_2/M^2$, this equation reads
\begin{equation}
0 = -v_{xx} - \frac{5}{x} v_x + \frac{6}{x^3} v + \frac{6\mu}{x^6}(3x + 2).
\label{Eq-KeyEq}
\end{equation}
Once we have solved this equation, the remaining amplitudes
$\dot{k}$ and $\dot{v}$ are obtained from (\ref{Eq-C1}) and
(\ref{Eq-C2}), respectively.

\subsection{The solutions of equation (\ref{Eq-KeyEq})}
A particular solution of (\ref{Eq-KeyEq}) is given by
\begin{displaymath}
v(x) = -\frac{2\mu}{3}\left( 1 - \frac{2}{x} +
\frac{3}{x^2} + \frac{3}{x^3} \right).
\end{displaymath}
In order to find the solutions of the homogeneous equation, one
performs the transformations $x = 24/z^2$, $v(x) = z^4 u(z)$,
which yields the Bessel differential equation
\begin{displaymath}
0 = z^2 u_{zz} + z u_z - (16 + z^2) u.
\end{displaymath}
The solutions are a linear combination of the Bessel functions
$J_4(iz)$ and $Y_4(iz)$. While $J_4(iz)$ behaves as $z^4$ for
small $|z|$, $Y_4(iz)$ has the expansion \cite{Walter}
\begin{eqnarray}
Y_4(iz) &=& -\frac{96}{\pi}\, z^{-4}\left( 1 - \frac{z^2}{12} +
\frac{z^4}{192}
            -\frac{z^6}{2304} + {\cal O}(z^8\log z)\right)
\nonumber\\
        &=& -\frac{1}{6\pi}\, x^2\left( 1 - \frac{2}{x} + \frac{3}{x^2}
            -\frac{6}{x^3} + {\cal O}(x^{-4}\log x)\right)
\label{Eq-Y4Exp}
\end{eqnarray}
near $z = 0$.
Thus, the general solution to (\ref{Eq-KeyEq}) is $v(x) = \mu \hat{v}(x)$, with
\begin{equation}
\hat{v}(x) =  -\frac{2}{3}\left( 1 - \frac{2}{x} +
\frac{3}{x^2} + \frac{3}{x^3} + \frac{C_1}{x^2} Y_4( i\sqrt{24/x} ) \right)
     + \frac{C_2}{x^2} J_4( i\sqrt{24/x} ) .
\label{Eq-vSol}
\end{equation}
While $v$ is regular at $x = \infty$ for any values of the constants
$C_1$, $C_2$, equation (\ref{Eq-C2}) shows that in order for $\dot{k}$
to be regular at $x = \infty$, it is necessary that $v$ decays at
least as fast as $x^{-2}$.  Comparing the expansion (\ref{Eq-Y4Exp})
with (\ref{Eq-vSol}) one sees that by choosing $C_1 = 6\pi$, one can
get rid of all terms which decay slower than $x^{-3}$. By looking at
gauge-invariant expressions, we will show later that this choice is
indeed necessary in order to get an asymptotic flat
solution. Therefore, $C_1$ is fixed by physical means.  The role of
the constant $C_2$ is discussed below.

\subsection{The Zerilli amplitudes}

In Ref. \cite{ST}, we have recently derived a gauge-invariant
generalization of the Zerilli equation which allows to study
perturbations on a Schwarzschild background written in any spherically
symmetric coordinates. In Appendix A of this paper we have written
the perturbed metric in terms of the generalized Zerilli function
$\psi $, that one obtains using the formalism of \cite{ST} for the
case at hand: $l=2$, even parity perturbations of a KS
background. This metric is a solution of Einstein's vacuum equations
provided the Zerilli function $\psi$ satisfies
\begin{equation}
\ddot{\psi } =
\frac{x-2}{x+2}\psi ^{''} + \frac{2}{x(x+2)} (\psi' - \dot{\psi }) +
\frac{4}{x+2}\dot{\psi '} -
\frac{6(3+6x+4x^2+4x^3)}{x^2(x+2)(3+2x)^2}\psi
\label{zerilli_equation}
\end{equation}
where $\psi = \psi(\tau, x)$, $\tau:= t/M$, and now $\dot{\psi } = \partial _{\tau} \psi $,
$\psi ' = \partial _x \psi $.

In order to evolve the KS
initial data, we have to relate the amplitudes $v$, $\dot{v}$ and $\dot{k}$
to the scalars $\psi$ and $\dot{\psi}$ (introduced in \cite{ST})
which satisfy the Zerilli equation (\ref{zerilli_equation}).
Using the expansions (\ref{Eq-Exp1},\ref{Eq-Exp2}) in
the expressions (\ref{Eq-KSgK1},\ref{Eq-KSgK2}),
it is straightforward to calculate $\psi$ and $\dot{\psi}$ using the
formulae given in \cite{ST}. The result is
\begin{eqnarray}
\psi &=& M \mu \left[ \frac{x^3\hat{v} - 6 }{3x(2x + 3)} \right] \, ,
\label{Eq-psi}\\
\dot{\psi} &=& -\mu\left[ \frac{ 2x^3(2x - 1) \hat{v} +
x^4(x - 2) \hat{v}' + 6(x-1) }{6x^2(2x+3)} \right]\, , \nonumber
\end{eqnarray}
where we have also used (\ref{Eq-C2}) in order to eliminate $\dot{k}$.
Since $\psi$ is gauge-invariant, it is clear that the free constants
$C_1$ and $C_2$ appearing in $v(x)$ cannot represent a gauge freedom.
In order to have an asymptotic flat solution, $v(x)$ must vanish at
infinity. As discussed in the previous subsection, this fixes the
value of the constant $C_1$. In this case, $\psi$ and $\dot{\psi}$
fall off like $x^{-2}$ at infinity. The constant $C_2$ is still free
and will determine different sets of initial data as one chooses its
value. The radiation content, as we will see, depends on $C_2$.  The
constant is therefore clearly associated with the ``spurious
radiation'' that the initial data contains with respect to
``astrophysically relevant'' initial data. One could probably
determine this content by evolving the initial data set backwards in
time. This calculation would be possible (at least for a limited
amount of time) within the confines of the close approximation
if the black holes are initially very close. One could therefore
follow the space-time backwards for a short time and see if 
incoming radiation is present at a finite distance of the holes.
We have not performed such a study, but it is feasible (we 
thank Jeff Winicour for bringing this to our attention).

It should be noticed that the initial data for the Zerilli function
diverges in the limit $x\rightarrow 0$ for all values of $C_2$, so one
cannot single out a preferred value of this constant by demanding the
initial data to be finite in this limit (even though, as already
mentioned, the multipole expansion is, in any case, valid only for $r
> \max\{a_1,-a_2\}$).

\subsection{The linearized apparent horizon equation}

Bishop {\em et al.} have argued that the position of the apparent
horizon is related to the constant $C_2$. Here we perform a linearized
analysis of the position of the horizon, to clarify the meaning of
their finding.

Given initial data $\bar{g}_{ij}$, $K_{ij}$ on a space-like slice $\Sigma$,
the location of an apparent horizon (AH) can be determined by the equation
\begin{equation}
\bar{\nabla}_i s^i - K_{ij} s^i s^j + K = 0,
\label{Eq-AHF}
\end{equation}
where $s^i$ is the unit outward normal to the AH.
If the AH is given by $f(\underline{x}) = 0$ for some function
$f$ on $\Sigma$, we have $s_i = \lambda \bar{\nabla}_i f$, where $\lambda^{-2}
= \bar{\nabla}^i f\bar{\nabla}_i f$.
For KS initial data of the form (\ref{Eq-KSgK1},\ref{Eq-KSgK2}),
it was assumed in \cite{BIMW} that the AH coincides with a
surface $S$ which is orthogonal to $k_i\,$. In this case, $s_ i =
-k_i/\alpha$, and one can check that equation (\ref{Eq-AHF})
is compatible with the result in \cite{BIMW}, i.e. $S$ is an AH if
$V = -1/2$ on $S$.

For spherically symmetric initial data,
\begin{eqnarray}
\bar{g}_{ij} dx^i dx^j &=& \gamma(r)^2 dr^2 + r^2 d\Omega^2,
\nonumber\\
K_{ij} dx^i dx^j &=& p(r)\gamma(r)^2 dr^2 + q(r) r^2 d\Omega^2\, ,
\nonumber
\end{eqnarray}
we must have $s_r = \gamma$, $s_A = 0$, and the AH equation (\ref{Eq-AHF})
yields $q = -1/(r\gamma)$. It is not difficult to show (either by
using the KS form of the Schwarzschild metric or more generally by integrating
the constraint equations) that this is equivalent to $r = 2M$, where
$M$ is the ADM mass.
We now want to linearize the AH equation around a spherically
symmetric background and to find the deviation of the AH from $r = 2M$.
In the linear regime, we expect that the location of the AH can be
described by the image of the circle $|\underline{x}| = 2M$
under a map of the form
\begin{displaymath}
\underline{x} \mapsto \underline{x} - \epsilon^2
D(\underline{x})\frac{\underline{x}}{r}\, .
\end{displaymath}
The deviation function $D(\underline{x})$ is related to the function
$f(\underline{x})$ as follows:
\begin{displaymath}
0 = f\left( \underline{x} - \epsilon^2 D(\underline{x})\frac{\underline{x}}{r} \right)
  = f^{(0)}(\underline{x}) + \epsilon^2\left[ \delta f(\underline{x}) - \partial_r f^{(0)}(\underline{x})
  D(\underline{x}) \right],
\end{displaymath}
where $f^{(0)}(\underline{x})$ is a function describing the AH
to zeroth order (for example $f^{(0)}(\underline{x}) = r - 2M$).
Using also $\lambda = \gamma(\partial_r f)^{-1}$, we obtain
$D(\underline{x}) = \lambda\delta f/\gamma$.

On the other hand, for linear perturbations around a spherically
symmetric background, the normalization of $s^i$ yields
$\delta s_r = \gamma h/2$ while equation (\ref{Eq-AHF}) gives
\begin{equation}
0 = \hat{\nabla}^A\left( \gamma\delta s_A - q_A \right) - r h
  + \frac{r^2}{2} k' + \gamma r^2 V_k\, ,
\end{equation}
where $h$, $k$, $q_A$ and $V_k$ are defined by
\begin{displaymath}
h = \gamma^{-2}\delta g_{rr}\, , \;\;\;
q_B = \delta g_{rB}\, , \;\;\;\
k = \bar{g}^{AB}\delta g_{AB}\, , \;\;\;
V_k = \delta(\bar{g}^{AB} K_{AB}) .
\end{displaymath}
In terms of the function $f$, $s_A = \delta(\lambda\partial_A f) = \gamma\partial_A D$,
and the linearized AH equation finally becomes
\begin{eqnarray}
\gamma^2\hat{\Delta} D = \hat{\nabla}^A q_A + r h - \frac{r^2}{2} k' - \gamma r^2 V_k\, .
\end{eqnarray}
Performing a multipolar decomposition of $D(\underline{x})$, this
equation becomes a set of algebraic equations for $D$.
Evaluating for the KS data proposed in Section II A, we find that
\begin{equation}
D(\underline{x}) = \alpha^2\left( \frac{\alpha^2}{3} ( 3r + 2M) v(r)
   - \frac{4M M_1 M_2}{r^2} \right) P_2(\cos\vartheta).
\end{equation}

So we see that the position of the apparent horizon
is given by the image of the circle $x = 2$ under the map
\begin{displaymath}
\underline{x} \mapsto \underline{x} -
\varepsilon^2\hat{D}(\underline{x})\,\frac{\underline{x}}{x}
\end{displaymath}
where the deviation function $\hat{D}(\underline{x})$ can be expressed
algebraically in terms of the perturbed three metric and extrinsic
curvature. For a KS metric and $x=2$,
\begin{displaymath}
\hat{D}(\underline{x}) = \frac{\mu}{2} \left( \frac{4}{3}\hat{v}(2) - 1 \right) P_2(\cos\vartheta).
\end{displaymath}
It is now clear that the deviation function depends on the value
of the constant $C_2\,$. In particular, we can choose $C_2$ such
that $\hat{D}(x=2)$ vanishes.

A way to see that the meaning of the constant $C_2$ is not just
a choice in the position of the apparent horizon is to notice that
once one has fixed the values of $M_1$ and $M_2$, the KS form of the
metric completely fixes the coordinates on the KS slice (at least to
linear order). Indeed, a gauge mode must satisfy the constraint
equations (\ref{Eq-C1},\ref{Eq-C2},\ref{Eq-C3}). On the other hand,
the general solution to these equations is completely determined by
$v(x)$, which cannot contain gauge modes since it is related to the
gauge-invariant amplitude $\psi$ according to (\ref{Eq-psi}).
Therefore, we have two possibilities to fix the coordinate location of
the apparent horizon (at $x=2$, say): The first possibility is to
perform an infinitesimal coordinate transformation such that the
apparent horizon appears unperturbed relative to the Schwarzschild
horizon.  In this case, the initial data will not have KS form
anymore.  The second possibility is to adjust the constant $C_2$ such
that the apparent horizon is at the location we desire.  Clearly,
these two methods are different since the former corresponds to a
gauge transformation, while the latter corresponds to a true physical
change in the initial data, as we anticipated before. So indeed the
constant $C_2$ is related to the position of the apparent horizon as
was noticed by Bishop {\em et al.}, but through a genuine change (not
just a gauge change) in the initial data.

\begin{figure}[h]
\centerline{\hbox{\psfig{file=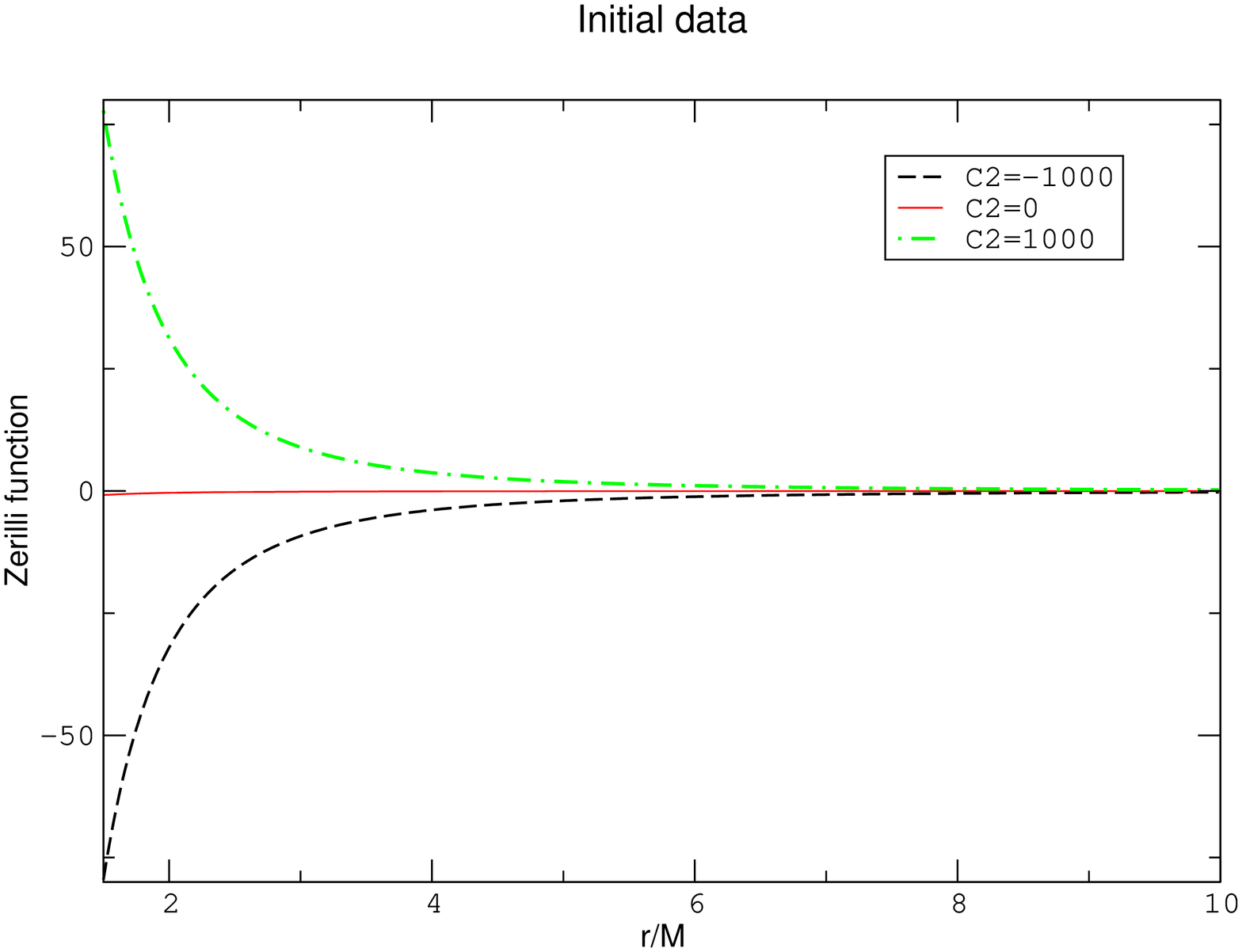,height=50mm}
\psfig{file=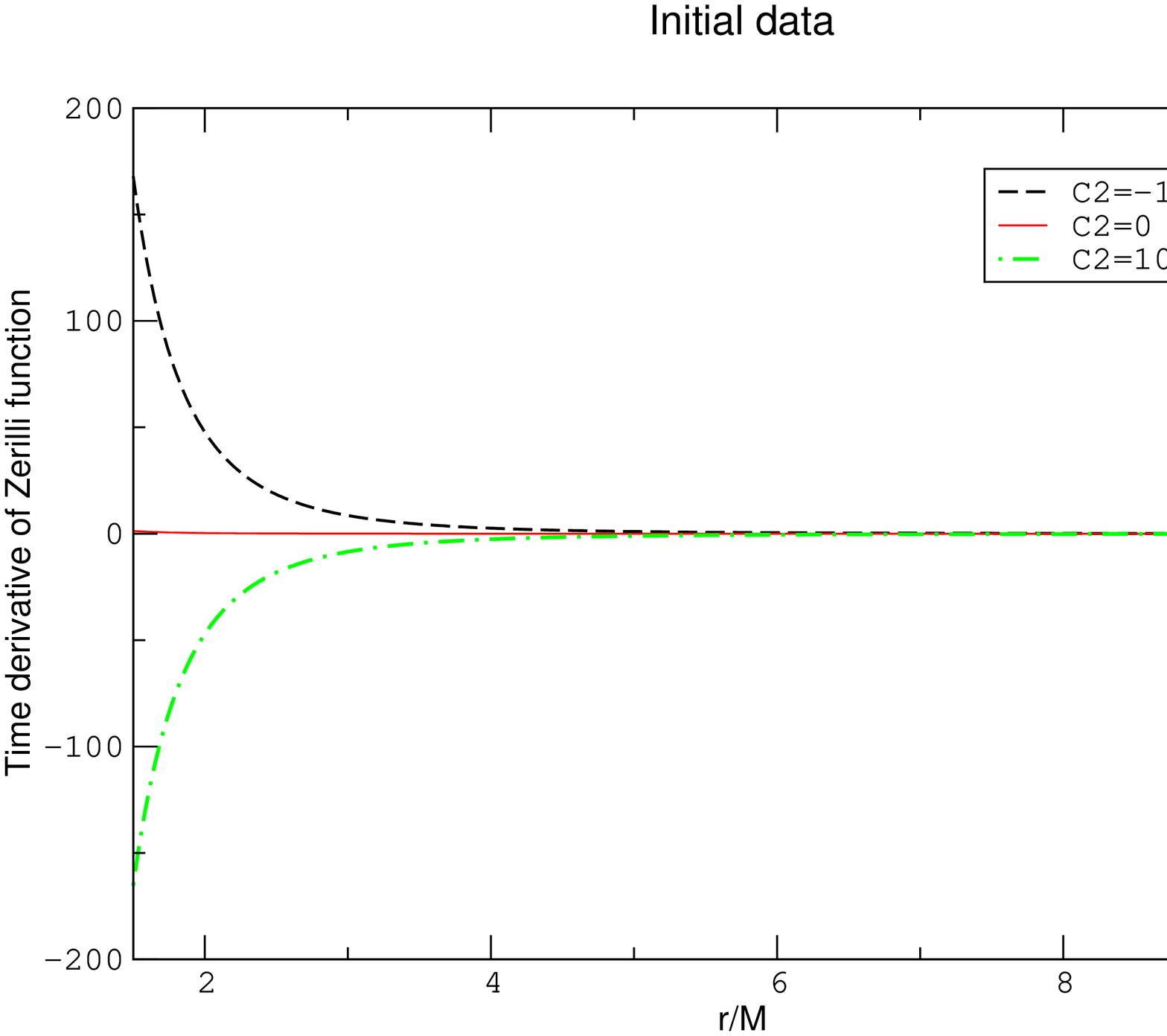,height=50mm}}}
\caption{The initial value of the Zerilli function and its time
derivative.  $C_2 = 0$ is nonvanishing, as may appear due to the
choice of scale of the figures.}
\end{figure}

Figures 1 and 2 show the initial data for the Zerilli function and its time
derivative for different values of $C_2$. In the next section we shall analyze
the dependence of the total radiated energy and waveforms on $C_2$.

\section{Evolution}

An expression for the radiated energy in terms of gauge-invariant
quantities is given in \cite{ST}. Since here we have expanded all
perturbations with respect to the Legendre polynomial
$P_2(\cos\vartheta) = \sqrt{4\pi/5}\, Y^{20}(\vartheta)$, this energy
expression becomes
\begin{displaymath}
\frac{dE}{du} = \frac{6}{5}\,\dot{\psi}^2 \; ,
\end{displaymath}
with $\dot{\psi }$ being evaluated in the radiative zone.

We have written a code that solves our generalized Regge-Wheeler and Zerilli
equations. As a consistency check, we have evolved the close limit of some
maximally sliced initial data (which can be seen as perturbations of
Schwarzschild in usual coordinates), being able to reproduce previous values
for the total radiated energy (e.g. Misner's initial data \cite{pp}, or
boosted black holes \cite{baker}).

The code is a standard second order dissipative, finite differencing,
one. In the case of a KS background we perform excision, i.e. we place
the inner boundary inside the black hole, and in that way avoid giving
boundary conditions there. In figure 3 we show the Zerilli function, scaled by $\mu$
(i.e. $\psi / \mu$) versus time, extracted at $r=100M$ for two different values of $C_2$.
{}From that plot one can notice, for example, the typical ringing
frequency for Schwarzschild black holes.

\begin{figure}[h]
\centerline{\psfig{file=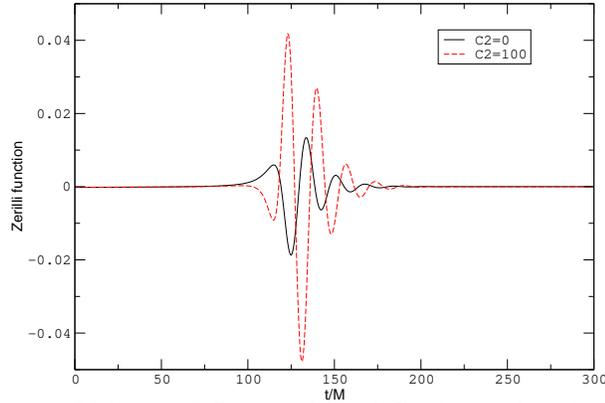,height=60mm}}
\caption{The radiated waveforms at $r=100M$ for two different values
of $C_2$. As usual in close limit collisions, the waveform is
dominated by the fundamental quasi-normal mode.}
\end{figure}

Given the linearity of Zerilli's equation, and the form of the KS
initial data, the dependence of the Zerilli function on the parameters
of the problem is
$$
\dot{\psi} (t,r) = \epsilon ^2 \mu \left(\dot{\psi}
_{a}(t,r) + C_2 \dot{\psi } _{b}(t,r)\right)
$$
where the functions $\psi _{a}$ and $\psi _{b}$ are
dimensionless. Accordingly, the total radiated energy is
$$
E = \frac{6 \epsilon ^4\mu ^2}{5}\left(\int
_0^{\infty}\dot{\psi_{a}}^2 dt + C_2 ^2 \int _0^{\infty} \dot{\psi
_{b}}^2dt + 2 C_2 \int _0^{\infty}\dot{\psi _{a}} \dot{\psi
_{b}}dt\right) $$
Therefore one needs to perform only three
runs to obtain the complete dependence of the radiated energy on the
free parameters. The result is
$$
E = \epsilon ^4 M \mu ^2 \left( 3.6 \times 10^{-4} +
5.2\times 10^{-7}C_2 ^2 - 2.2
\times 10^{-5}C_2  \right)
$$
The quantity $E/(\epsilon ^4 M \mu ^2)$, as a (quadratic) function of $C_2$,
has a local minimum at $C_2 \approx 21$, where it takes the value
$E/(\epsilon ^4 M \mu ^2) \approx 1.3 \times 10^{-4}$. This function is plotted
in figure 3.

\begin{figure}[h]
\centerline{\psfig{file=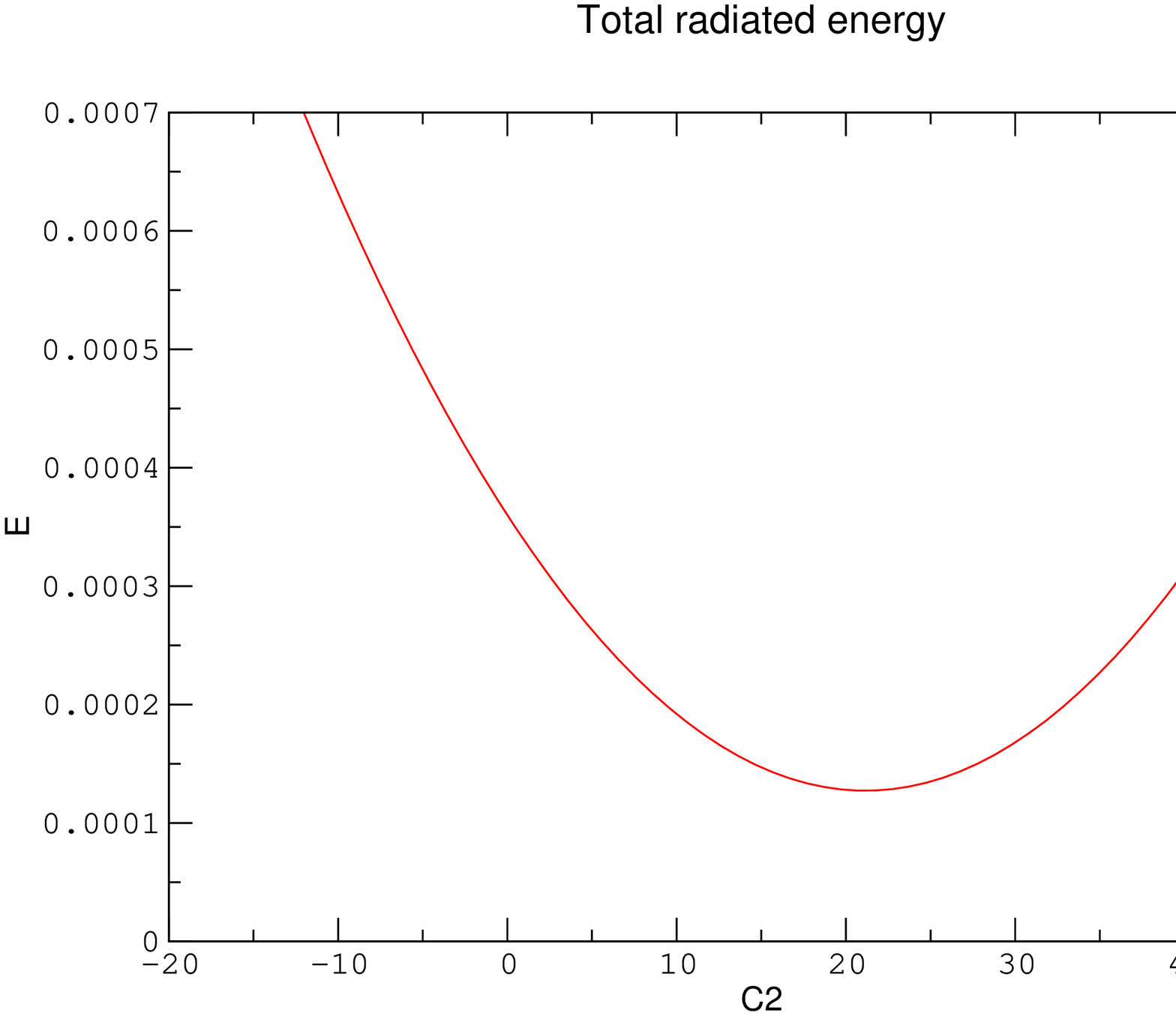,height=60mm}}
\caption{The radiated energy (in units of $\epsilon^4 M\mu^2$) as a
function of $C_2 $. As a rough comparison, if we consider equal mass
black holes ($\mu=1$), and take the ``separation'' $\epsilon=1$ and
identify it with the ``separation in the conformal background
geometry'' for two Brill--Lindquist black holes (for a more physical
picture, a conformal separation of less than $0.8M$ corresponds to a
common apparent horizon for the Brill--Lindquist family), the latter
would radiate around $10^{-5}M$, which is rougly similar to the
radiation we get for the minimum value of $C_2$.}
\end{figure}

\section{Discussion}

We have evolved the initial value family of Bishop {\em et al.} in the
limit in which the black holes are close to each other by treating the
spacetime as a single distorted Kerr--Schild black hole and solving
the linearized Einstein equations for the distortion. The evolution
sheds further light on the role of the integration constants present
in the family.

An obvious question to ask would be ``does this family contain
more/less radiation than other families'' (for instance the Misner
data). Unfortunately, the family has explicit free parameters and
therefore the comparison is highly dependent on the arbitrary values
of these parameters. This should not be misinterpreted as a problem:
it just highlights that the initial value problem for binary black
holes inevitably contains ambiguities. Some proposals may resolve the
ambiguities based on aesthetic criteria, but from a physical point of
view that is not more satisfactory than simply picking values for the
constants involved.  These issues could be better understood if one
evolved the systems backwards in time and tried to establish the
amount of incoming radiation. The present results should be of
interest in the calibration of numerical codes based on the
Kerr-Schild coordinate system. Experiments with the Maya binary black
hole code \cite{maya} to compare results are currently under way.

\begin{acknowledgments}
This work was supported in part by the Swiss National Science Foundation, by
 grants NSF-PHY-0090091, NSF-PHY-9800973, by Fundaci\'on Antorchas,
by the Eberly Family Research Fund at Penn State and the Horace C.
Hearne  Jr. Institute of Theoretical Physics.
We wish to thank Jeff Winicour for comments and for helping us clarify
an earlier version of the paper.
After completing this work we learnt that Carsten K\"ollein at Albert
 Einstein Institute performed a perturbative evolution of this same
 family of initial data in his M.Sc. thesis (unpublished) using the
Teukolsky equation. We thank Manuela Campanelli for bringing this to
our attention.
\end{acknowledgments}

\appendix
\section{Perturbed metric for a KS background}
For the particular case in which the background metric is KS, the
perturbed $l=2$ even metric in the Regge--Wheeler gauge, in terms of
the Zerilli function, is
\begin{eqnarray*}
g_{rr} &=&  1 + \frac{2}{x} + \frac{\delta }{M}\left[
\frac{6(x+2)(3+6x+4x^2+4x^3)}{(3+2x)^2x^3}\psi +
\frac{4(4x^2+9x+3)}{x^2(3+2x)}\dot{\psi } -    \right. \\
& & \left. \frac{2(2x^3+2x^2+15x+6)}{x^2(3+2x)}\psi ' - 2x\psi ^{''}
\right]Y^{20}(\vartheta) \\
& &              \\
g_{rt } &=& \frac{2}{x} + \frac{\delta }{M} \left[
\frac{12(3+6x+4x^2+4x^3)}{x^3(3+2x)^2}\psi +
\frac{4(2x^2-6x-3)}{x^2(3+2x)}\psi ' -  \right.  \\
& & \left. \frac{2(x+2)(2x^2-3-6x)}{x^2(3+2x)}\dot{\psi } -
  2x\dot{\psi '} \right ]Y^{20}(\vartheta) \\
& & \\
g_{\theta \theta } &=& x^2 + \frac{\delta }{M} \left[-\frac{12(x+1+x^2)}{3+2x}\psi -
4x\dot{\psi } - 2x(x-2)\psi ' \right]Y^{20}(\vartheta)            \\
& & \\
g_{\phi \phi } &=& g_{\theta \theta } \sin ^2{\theta } \\
& & \\
g_{tt} &=&  -1 + \frac{2}{x} + \frac{\delta }{M}\left[
\frac{6(4+x^2)(3+6x+4x^2+4x^3)}{(3+2x)^2x^3(x+2)}\psi +
\frac{4(5x^2+15x+6)}{x^2(3+2x)(x+2)}\dot{\psi } -    \right. \\
& & \left. \frac{2(2x^4-2x^3-5x^2+24x+12)}{x^2(3+2x)(x+2)}\psi ' -
2x\frac{x-2}{x+2}\psi^{''} - \frac{8x}{x+2} \dot{\psi '}\right]Y^{20}(\vartheta)
\\
\end{eqnarray*}
where $\delta $ is a perturbative parameter.


\end{document}